\documentclass[12pt]{article}
\usepackage[utf8]{inputenc}
\usepackage{authblk}
\usepackage{setspace}
\usepackage[margin=1.25in]{geometry}
\usepackage{graphicx}
\usepackage{subcaption}
\usepackage{amsmath}
\usepackage{float}

\usepackage[backend=bibtex,style=nejm, 
citestyle=numeric-comp,
sorting=none]{biblatex}

\begin{document}
\title{Investigating the Two-Dimensional Generalized XY Model using Tensor Networks}
\author[1*]{Vamika Longia}
\author[1,2]{Anosh Joseph}
\author[3]{Abhishek Samlodia}
\affil[1]{Department of Physical Sciences, Indian Institute of Science Education and Research - Mohali, Knowledge City, Sector 81, SAS Nagar, Punjab 140306, India}
\affil[2]{National Institute for Theoretical and Computational Sciences, School of Physics and Mandelstam Institute for Theoretical Physics, University of the Witwatersrand, Johannesburg, Wits 2050, South Africa}
\affil[3]{Department of Physics, Syracuse University, Syracuse, New York 13244, United States}
\affil[*]{vamika.m.longia@gmail.com}

\onehalfspacing
\date{}
\maketitle

\begin{abstract}

The critical behavior of the two-dimensional XY model has been explored in the literature using various methods. They include the high-temperature expansion (HTE) method, Monte Carlo (MC) approach, strong coupling expansion method, and tensor network (TN) methods. This model undergoes a Berezinskii-Kosterlitz-Thouless (BKT) type of phase transition. This model can be modified by adding spin-nematic interaction terms with a period to give rise to the generalized XY model. The modified model contains excitations of integer and half-integer vortices. These vortices govern the critical behavior of the theory and produce rich physics. With the help of tensor networks, we investigate the transition behavior between the integer vortex binding and half-integer vortex binding phases of the model and how this transition line merges into two BKT transition lines.

\end{abstract}

\section{The Model}

The model we study, the two-dimensional generalized XY model on a square lattice, has the following Hamiltonian
\begin{equation}
H = - \left( J \sum_{\langle ij \rangle} \cos(\theta_i - \theta_j) \;\; + \;\; J_1 \sum_{\langle ij \rangle} \cos(q (\theta_i - \theta_j)) \;\; + \;\; h \sum_i \cos \theta_i \right),
\end{equation}
where $\theta_i \in [0, 2\pi]$ is the spin orientation at the lattice site $i$, $\cos(q (\theta_i - \theta_j))$, with $q > 1$, is the nematic term in the model, $J > 0$ and $J_1 > 0$ are the coupling constants between the nearest-neighbor spins, $h$ is the applied external magnetic field which breaks the global $O(2)$ symmetry, $\langle ij \rangle$ denotes the summation over the nearest-neighbor sites $i$ and $j$ for the square lattice. 

For $q = 2$, $J = \Delta$, $J_1 = (1 - \Delta)$, and $h = 0$ the Hamiltonian takes the form \cite{Lach_2020}
\begin{equation}
\label{eq:delta_hamiltonian}
    H = - \left( \sum_{\langle i j \rangle} \Delta \cos(\theta_i - \theta_j) \; \; + \; \;  \sum_{\langle i j \rangle} (1 - \Delta) \cos(2(\theta_i - \theta_j)) \right).
\end{equation}

The usual XY spin model with integer vortex excitations can be obtained from Eq. \eqref{eq:delta_hamiltonian} by setting $\Delta = 1$. By setting $\Delta = 0$ in Eq. \eqref{eq:delta_hamiltonian} we can study a purely spin-nematic model possessing symmetry under the $\theta_i \to \theta_i + \pi$ transformation.

We study the model in the regime, $0 \leq \Delta \leq 1$, where the system undergoes two phase transitions: Ising-like, which exhibits continuous phase transition, and BKT class, which exhibits infinite order phase transition, with a multi-critical point. The three phases are the integer vortex pair phase,  the half-integer vortex pair phase, and a disordered phase~\cite{Song_2021}. After that, we study the model in the presence of non-zero $h$.

\section{Tensor Network Construction}

\subsection{The Partition Function}

The partition function is given by
\begin{equation}
Z = \prod_i \int \frac{d\theta_i}{2\pi} \prod_{ \langle ij \rangle} e^{\beta [\Delta \cos(\theta_i - \theta_j) + (1 - \Delta) \cos(2(\theta_i - \theta_j))]} \prod_i e^{\beta h \cos(\theta_i)}.
\end{equation}

Using the character expansion of the Boltzmann factor to construct the tensor network representation where the phase variable changes into number indices on the links~\cite{Yu:2013sbi}. This transformation, in terms of modified Bessel functions of the first kind, $I_n(x)$ is
\begin{equation}
e^{x \cos\theta} = \sum_{n = -\infty}^\infty I_n(x)e^{i n \theta}.
\end{equation}

Using the above expansion, the partition function takes the form
\begin{equation}
\label{eq:Z_transformed} 
Z = \prod_s \int \frac{d\theta_s}{2\pi} \prod_{l \in L} \sum_{n_l} a_{n_l}(\beta, \Delta) e^{i n_l(\theta_{s_i} - \theta_{s_j})} \times \sum_{p_l} I_{p_{l}}(\beta h) e^{i p_l \theta_{s_i}}.
\end{equation}
In the above $l$ runs over all the links, $s$ labels all the lattice sites, and 
\begin{equation}
\label{eq:Z_transformed_an}
    a_n (\beta, \Delta) = \sum_{m = -\infty}^\infty I_{n - 2m}(\beta\Delta) I_m(\beta(1 - \Delta)).
\end{equation}

After integration over the $\theta$ variables, we obtain the tensor form of the partition function
\begin{equation}
\label{eq:Z}
Z = {\rm tTr} \left(\prod_s T_{n_1,n_2}^{n_3,n_4}(s) \right),
\end{equation}
with
\begin{equation}
\label{eq:Z_tensor}
T_{n_1,n_2,n_3,n_4}(s) = \left( \prod_{i=1}^4a_{n_i}(\beta,\Delta) \right)^{\frac{1}{2}} ~I_{n_1+n_2-n_3-n_4}(\beta h),
\end{equation}
and ${\rm tTr}$ indicates the tensor trace. 

The indices $(n_1$, $n_2$, $n_3$, $n_4)$ $\in$ $(-\infty$, $\infty)$ denote the four legs of the tensor. For computational purposes, the size of each leg is truncated down to a preferred bond dimension $D$. Similarly, the index $m$ in Eq. \eqref{eq:Z_transformed_an} is truncated to a fixed number large enough for convergent results. Furthermore, for $h = 0$ the second term in Eq. \eqref{eq:Z_tensor} becomes $\delta_{n_1, n_2}^{n_3, n_4}$~\cite{Jha:2020oik}.

\subsection{Observables}

We use the following set of observables to find the transition temperature in the model
\begin{equation}
F = - \frac{1}{V}\frac{\ln Z}{\beta},  \;\; C_{_V} = \frac{\partial^2 F}{\partial \beta^2}, \;\; M = -\frac{\partial F}{\partial h} = \frac{1}{\beta} \frac{\partial \ln Z}{\partial h}, \;\;  \chi = \frac{\partial M}{\partial h} \; = \; \frac{1}{\beta}\frac{\partial^2 \ln Z}{\partial h^2},
\end{equation}
where, $F$ is the free-energy density, $C_{_V}$ is the specific heat, $M$ is the magnetization, and $\chi$ is the magnetic susceptibility.

Using the partition function given in Eq. \eqref{eq:Z}, the tensor formulation of $M$ can be expressed as~\cite{Jha:2020oik}
\begin{equation}
M = {\rm tTr} \left( \sqrt{\prod_{i = 1}^4 a_{n_i}(\beta, \Delta)} \times \frac{I_{n_1 + n_2 - n_3 - n_4 - 1}(\beta h) + I_{n_1 + n_2 - n_3 - n_4 + 1}(\beta h)}{2} \right).
\end{equation}

\section{Numerical Simulations}

We use the Higher Order Tensor Renormalization Group (HOTRG) method for the coarse-graining step of the tensor, as described in \cite{PhysRevB.86.045139, Butt:2022qqx}. This step is repeated for a fixed number of iterations to converge the simulations. Then, in the end, we compute the final partition function and other observables, such as free-energy density and magnetization. We first study the case where the external magnetic field is set to zero and find the critical temperature by locating the specific heat peak for a fixed $\Delta$. Since, there are both types of phase transitions of which the BKT being infinite-order has the continuous derivatives of $\log Z$, the specific heat is not enough to precisely locate the critical points. Hence, we compute the peak of magnetic susceptibility with respect to temperature for various values of $h$ to obtain a much more accurate transition.

\section{Summary}

We have reported on how to investigate the phase diagram of the two-dimensional generalized XY model for $q = 2$ in $\cos(q(\theta_i-\theta_j))$ using the HOTRG method. We can use the specific heat and magnetic susceptibility to locate the critical points of the transitions belonging to different types - Ising for the first and BKT for the second. 

We use a GPU-accelerated~\cite{Jha:2023bpn} version of the HOTRG method to reduce the simulation time considerably. Our initial results are consistent with the conventional Monte Carlo studies~\cite{PhysRevE.87.062112} but with greater precision. We will soon report on our detailed simulation results elsewhere. 

\subsection*{Acknowledgements}

We thank Raghav Jha for ongoing discussions and collaboration on this work. Numerical calculations were carried out at IISER Mohali and the PARAM SMRITI supercomputing facility at the National Agri-Food Biotechnology Institute - Center of Innovative and Applied Bioprocessing (NABI-CIAB). We acknowledge the National Supercomputing Mission (NSM) for providing computing resources of PARAM SMRITI at NABI, Mohali, which is implemented by C-DAC and supported by the Ministry of Electronics and Information Technology (MeitY) and Department of Science and Technology (DST), Government of India. Part of the simulations were carried out at Condor, Syracuse University Cluster as well.

\printbibliography

\end{document}